\documentclass[epj,referee]{svjour}
\newcommand{\tfrac}[2]{{\textstyle\frac{#1}{#2}}}
\usepackage{graphics,epsf}
\begin{document}

\title{Vertical Melting of a Stack of Membranes} 

\author{M.E.S. Borelli \thanks{\email{borelli@physik.fu-berlin.de}}
\and H. Kleinert \thanks{\email{kleinert@physik.fu-berlin.de}} \and
Adriaan M.J. Schakel\thanks{\email{schakel@physik.fu-berlin.de}}}

\institute{Institut f\"ur Theoretische Physik, Freie Universit\"at
Berlin, Arnimallee 14, 14195 Berlin, Germany}

\date{\today}

\abstract{A stack of tensionless membranes with nonlinear curvature
energy and vertical harmonic interaction is studied.  At low
temperatures, the system forms a lamellar phase. At a critical
temperature, the stack disorders vertically in a melting-like
transition.
\PACS{{82.65.Dp}{Thermodynamics of surfaces and
interfaces} \and {68.35.Rh}{Phase transitions and critical phenomena}}
} \maketitle

\section{Introduction} \label{intro}

Under suitable conditions, lipid membranes in aqueous solution are
known to form lamellar structures, characteri\-zed by a parallel
arrangement of the membranes alter\-nating with thin layers of water
\cite{helfstack}.  The existence of such structures is in contrast to
the behavior of a single tensionless membrane subject to thermal
fluctuations, which is always in a disordered, crumpled phase, filling
the embedding space completely \cite{Jerusalem}.  In a stack, this
phase is suppressed by the steric repulsion between the membranes
which prevents them from passing through each other \cite{helfstack},
thus constraining the amplitude of the height fluctuations of each
membrane to be less than the distance to its nearest neighbors.

In this note, we investigate the question whether the lamellar
structure exists at all temperatures, or whether thermal fluctuations
can destroy the vertical order at some critical temperature.

Such a transition does not take place in the simplest model of a stack
proposed by Helfrich \cite{helfstack}, where the membranes possess
only a linearized curvature energy and a harmonic repulsive term
accounts for the vertical forces in the stack, approximating in a
rough way the steric repulsion.  In this purely Gaussian
approximation, the theory is equivalent to de Gennes' theory of
smectic-A liquid crystals, having only an ordered phase
\cite{DGstack}.

In this paper we extend the simplest model by taking into account
higher-order terms of the curvature energy, and show that thermal
fluctuations cause a finite stack of membranes to disorder vertically,
a process which we may call vertical melting.  A similar phase was
previously described in a different context by Huse and Leibler
\cite{HuLe}, and by Kleinert \cite{nematic}, who related the molten
phase to the directionally ordered phase of a nematic li\-quid
crystal. We shall find that thermal fluctuations renormalize both the
coefficient of the curvature term and the vertical compressibility of
the stack.  Analyzing the renormalization group equations, we find,
besides the Gaussian fixed point governing the low-temperature phase,
a nontrivial fixed point determining the critical exponents of the
vertical melting transition.

\section{The Model} \label{model}

We consider a generalization of the Helfrich model due to Janke and
Kleinert \cite{JankeKleinert}, in which a multilayer system is made up
of $(N+1)$ fluid membranes, parallel to the $xy$ plane of a Cartesian
coordinate system, separated a distance $l$.  If the vertical
displacement of the $m$th membrane with respect to this reference
plane is described by a function $u_m({\bf x}) \equiv u({\bf
x}_\perp,m l)$, where ${\bf x}_\perp = (x,y)$, the energy of the stack
reads:
\begin{equation} \label{mod0}
E = \sum_{m} \int {\rm d}^2 x_\perp \sqrt{g_m} \left[r_0 + \tfrac{1}{2}
\kappa_0 H_m^2 +  \frac{B_0}{2 l} (u_m - u_{m-1})^2 \right].
\end{equation}
Here, $H_m= \partial_i N_{m,i}$ is the mean curvature, where ${\bf
N}_m \propto (-\partial_1 u_m, -\partial_2 u_m, 1)$ is the unit normal
to the $m$th membrane, and $g_{m,i j}=\delta_{i j} + \partial_i u_m
\partial_j u_m$ the induced me\-tric, with $i,j = 1,2$, $\partial_1 =
\partial/\partial x, \partial_2 = \partial/\partial y$ and $g_m = \det
[g_{m,i j}]$. The parameter $r_0$ is the surface tension of a single
membrane, $\kappa_0$ its ben\-ding rigidity, and $B_0$ the compressibility
of the stack.  In Eq.\ (\ref{mod0}), as in the following, the subscript
$0$ denotes bare quantities, whereas renormalized parameters will carry
no subscript.

In the original Helfrich model, the surface tension $r_0$ was not
included because the membranes in the stack are tensionless.  We have
included $r_0$ in the energy (\ref{mod0}) as an infrared regulator
which also serves to absorb ultraviolet infinities.  After carrying
out the various integrals, the renormalized physical tension will be
set equal to zero.

For slow spatial variations, the discrete variable $m l$ may be replaced
with a continuous one, and $u({\bf x}_\perp,m l) \to u({\bf x})$, where
${\bf x} = ({\bf x}_\perp,z)$.  In this limit, the ener\-gy
(\ref{mod0}) reduces to
\begin{equation} \label{mod1}
E = \int^{L_{\parallel}}_0 
{\rm d}z \int {\rm d}^2 x_\perp \sqrt{g} \left[ \sigma_0 +
\tfrac{1}{2} K _0 H^2 + \tfrac{1}{2} B_0 (\partial_z u)^2 \right].
\end{equation}
Here we have introduced bulk versions of the surface tension $\sigma_0
\equiv r_0/l$ and ben\-ding rigidity $K_0 \equiv \kappa_0/l$, and
defined $L_{\parallel} \equiv N l$.  In Ref.\ \cite{nematic} the
vertical gradient energy $(\partial_z u)^2$ was replaced by the normal
gradient energy $({\bf N} \cdot \nabla u)^2$, which is physically more
correct and has the advantage of being reparametrization invariant
(see also Ref. \cite{foltin}).
In the following, we shall derive all results for both terms and
analyze the difference between the two.

We study the stack perturbatively, starting with the lamellar
configuration, and expand the theory in the inverse parameter
$\alpha_0 = 1/K_0$, which is assumed to be small.  To keep our notation
in conformity with the li\-terature, we use $K_0$ throughout the text
and resort to its inverse only when necessary.

\section{Renormalization} \label{renormalization}

The phase transition to be described in this note is caused by a
competition between the sof\-tening of the ben\-ding rigidity due to
thermal fluctuations and the stack-preser\-ving vertical elastic
forces.  To understand this competition we study how thermal
fluctuations renormalize the parameters of the theory.  For this we
expand the energy (\ref{mod1}) up to fourth order in the displacement
field $u({\bf x})$, where it reads
\begin{eqnarray} \label{ourmodel}
\! E = \int^{L_{\parallel}}_0 \! \! \! \! 
{\rm d}z \! \! \int \! {\rm d}^2 x_\perp \Bigl[ &\tfrac{1}{2}&
\sigma_0 (\partial_i u)^2 + \tfrac{1}{2} K_0 (\partial_i^2 u)^2 +
\tfrac{1}{2} B_0 (\partial_z u)^2 \nonumber \\ &-& \tfrac{1}{8}
\sigma_0 (\partial_i u)^2 (\partial_j u)^2 - \tfrac{1}{4} K_0
(\partial_i^2 u)^2 (\partial_j u)^2 \nonumber \\ &-& K_0 (\partial_i
u)(\partial_j u) (\partial_i \partial_j u) (\partial_k^2 u)\nonumber
\\ &\pm& \tfrac{1}{4} B_0 (\partial_z u)^2 (\partial_i u)^2 \Bigr],
\end{eqnarray}
where the lower sign in the last term refers to the more physical
normal gradient energy $({\bf N} \cdot \nabla u)^2$.
For zero surface tension, the lowest-order contribution to the energy
due to longitudinal displacements is of the usual elastic form
$\tfrac{1}{2} B_0 (\partial_z u)^2$, while that due to transverse
displacements is of higher order, viz.\ $\tfrac{1}{2} K_0 (\partial_i^2
u)^2$.  

The one-loop contributions can be calculated either by using Feynman
diagrams, or by a derivative expansion, as in Ref.\ \cite{memb}.
Following standard procedure, we integrate out fluctuations with
transverse wavevectors in a momentum shell.  Since the stack is
periodic and of finite extent in the $z$-direction, the Fourier
transform includes a sum $(1/L_{\parallel})\sum_{n=-N/2}^{N/2}$ over
the discrete wavevector components
\begin{equation} \label{matsubara}
\omega_n = \frac{2 \pi}{L_{\parallel}} n.
\end{equation}
We take into account the interlayer spacing $l$ in a rough way by
restricting the values of the discrete variables to $|\omega_n| <
\pi/l$, so that the summation index $n$ lies in the interval
$-\tfrac{1}{2}N<n<\frac{1}{2}N$.  To one-loop order, the bare
parameters are renormalized to
\begin{equation} 
\sigma_0 \to \sigma_0(1 + I_1), \;\;\; K_0 \to K_0 (1 - \tfrac{3}{2} I_2),
\;\;\; B_0 \to B_0 (1 \pm \tfrac{1}{2} I_2 ),
\end{equation} 
where 
\begin{eqnarray}
I_1 &=& \frac{k_{\rm B}T}{L_{\parallel}}\sum_{n=-N/2}^{N/2} \int \frac{{\rm
d}^2q_\perp}{(2\pi)^2} \frac{\tfrac{1}{2} \frac{B_0}{\sigma_0} \omega_n^2 -
q_\perp^2 - \tfrac{3}{2} \frac{K_0}{\sigma_0} q_\perp^4}{B_0 \omega_n^2 +
\sigma_0 q_\perp^2 + K_0 q_\perp^4}  \label{i1} \\ I_2 &=&
\frac{k_{\rm B}T}{L_{\parallel}}\sum_{n=-N/2}^{N/2} \int \frac{{\rm
d}^2q_\perp}{(2\pi)^2} \frac{q_\perp^2}{B_0 \omega_n^2 + \sigma_0
q_\perp^2 + K_0 q_\perp^4}. \label{i12}
\end{eqnarray}
We regularize the integrals in the ultraviolet by introdu\-cing a
sharp transverse wavevector cutoff $\Lambda$ inversely proportional to
the lateral size $a$ of a lipid molecule.  Actually, the divergent
contributions to the above integrals are independent of $B_0$. This is
due to the discrete nature of the stack.  By restricting the values of
the discrete wavevectors $\omega_n$ to account for the interlayer
spacing, as explained above, all divergences proportional to $B_0$ are
suppressed.

The renormalization flow is obtained by integrating out transverse
wavevectors in a momentum shell $\Lambda/s < q_\perp < \Lambda$, and
subsequently rescaling the coordinates.  We thus obtain,
\begin{equation} \label{I2}
I_1 = \tfrac{1}{2} I_2 = \frac{1}{4 \pi} \frac{k_{\rm B}T}{L_{\parallel}}
\frac{1}{K_0} (N + 1) \ln s.
\end{equation}

Since these results are independent of $\sigma_0$, we can safely set
the surface tension to zero, thus describing a stack of tensionless
membranes, characterized by the two remaining parameters, $B$ and $K$.
Under a rescaling ${\bf x}_\perp \to {\bf x}_\perp/s$ of the
coordinates in the plane and $z \to z/s^{z_{\rm c}}$ along the stack
axis, the expansion parameter of perturbation theory $\alpha \equiv
1/K$ scales like $\alpha \to s^{-z_{\rm c}} \alpha$, and the
compressibility scales like $B \to s^{4 - z_{\rm c}} B$. Here, $z_{\rm
c}$ allows for the possibility of anisotropic scaling.  From Eq.\
(\ref{ourmodel}) with $\sigma=0$, it follows that $z_{\rm c}=2$ in the
Helfrich model.  Using the above results, one readily generates
differential recursion relations to lowest nontri\-vial order
\begin{eqnarray}
 \frac{{\rm d} \alpha}{{\rm d} \ln s } &=& - z_{\rm
c} \alpha + \frac{3}{4 \pi} \frac{k_{\rm B}T}{L_{\parallel}} \alpha^2 (N + 1)
\label{betaalpha} \\  \frac{{\rm d} B}{{\rm d} \ln s }
 &=& (4 - z_{\rm c}) B \pm \frac{1}{4 \pi} \frac{k_{\rm
B}T}{L_{\parallel}} B \alpha (N + 1). \label{betanu}
\end{eqnarray}
Besides the Gaussian fixed point $(\alpha=0, B=0)$ which is stable in
the infrared, the flow equations also admit a nontrivial, unstable fixed
point at
\begin{equation} \label{fp}
\alpha^*= \frac{4 \pi}{3} \frac{L_{\parallel}}{k_{\rm B}T} \frac{z_{\rm c}}{N+1},
\;\;\; B =0.
\end{equation} 
The latter implies the presence of a phase transition at a critical
temperature 
\begin{equation} \label{tc}
k_{\rm B} T_{\rm c} = \frac{4 \pi}{3} \frac{N}{N+1} \frac{l z_{\rm
c}}{\alpha^*} .
\end{equation}

\begin{figure}[h]
\begin{center}
\epsfxsize=8.cm \mbox{\epsfbox{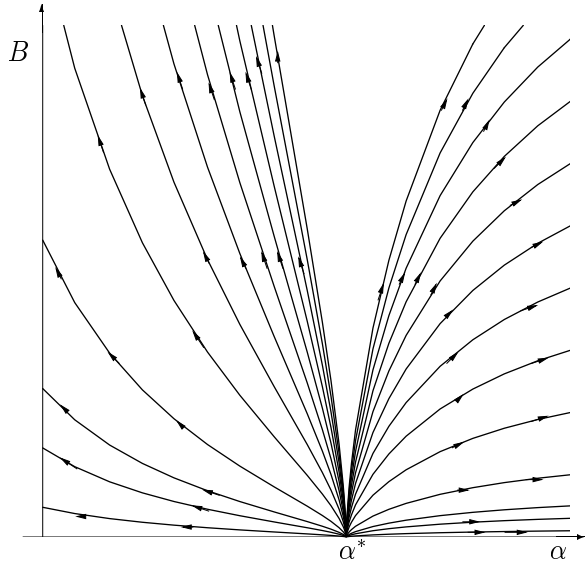}}
\end{center}
\caption{Flow diagram in the $(\alpha,B)$-plane. The diagram is
plotted using the lower sign in the last term in Eq.\ (\ref{betanu}).
\label{fig:flow}}
\end{figure}

The flow diagram corresponding to the above system of differential
equations is shown in Fig.\ \ref{fig:flow}.
For $T<T_{\rm c}$, the inverse ben\-ding rigidity $\alpha$ flows to
the Gaussian fixed point at the origin.  In this low-temperature
phase, the ben\-ding rigidity of the membranes increases with their
lateral size, and thermal fluctuations are suppressed.  This
weak-coupling phase is the lamellar pha\-se, where the rotational
symmetry is spontaneously broken.  For $T>T_{\rm c}$, on the other
hand, $\alpha$ flows with increasing length scales away from the
nontrivial fixed point at $\alpha^*$ in the opposite direction.  As
$\alpha$ increases, the ben\-ding rigi\-dity decreases and the
membrane fluctuations become stronger.  At the critical point
$\alpha^*$, the stack disorders vertically and the system enters a
strong-coupling disordered phase.  Note that the critical
tempe\-rature (\ref{tc}) depends only weakly on the number $(N+1)$ of
membranes.

The flow equations (\ref{betaalpha}) and (\ref{betanu}) can be
integrated exactly, yielding:
\begin{equation}
B = c \, \alpha^{\pm 1/3} \left| \frac{\alpha -
\alpha^*}{\alpha}\right|^{4/z_{\rm c} - 1 \pm 1/3}, \label{B}
\end{equation}
where $c$ is an integration constant. For $z_{\rm c}=2$, the exponent
is equal to $1 \pm 1/3$.  Explicitly, as we approach $T_{\rm c}$ from
below, $B$ goes to zero as
\begin{equation}
B \sim |T - T_{\rm c}|^{1 \pm 1/3}.
\end{equation}

\begin{figure}[h]
\begin{center}
\epsfxsize=8.cm \mbox{\epsfbox{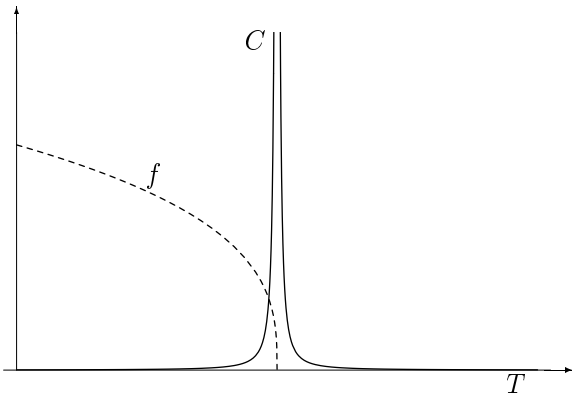}}
\end{center}
\caption{Free energy density and specific heat of a stack of
membranes, with the lower sign in Eqs.\ (\ref{freeplot}) and
(\ref{cplot}). \label{fig:free}}
\end{figure}

The free energy density of the model in the lamellar phase can be calculated
in the harmonic approximation, as in Ref.\ \cite{helfstack}.  For a
finite stack, it reads
\begin{equation}
f = \frac{1}{16 \pi} \frac{k_{\rm
B}T}{L_{\parallel}^2} \left(\frac{B}{K}\right)^{1/2}N(N + 2)
\end{equation}
Thus, as $T$ approaches $T_{\rm c}$ from below, the free energy density
behaves for $z_{\rm c}=2$ like
\begin{equation}
  \label{freeplot}
f \sim |T - T_{\rm c}|^{1/2 \pm 1/6},
\end{equation}
and the specific heat of the stack diverges as
\begin{equation}
  \label{cplot}
C \sim |T - T_{\rm c}|^{-3/2 \pm 1/6}.
\end{equation}
Figure \ref{fig:free} shows a plot of the free energy density and
specific heat of the stack for the lower sign in Eqs.\
(\ref{freeplot}) and (\ref{cplot}). 

We thus see that by using either the vertical gradient energy
$(\partial_z u)^2$ or the more physical normal gradient energy $({\bf
N}\cdot \nabla u)^2$, the qualitative behavior of the stack of
membranes is not altered, but the critical exponents of the melting
transition differ from each other.  This is due to the fact that the
normal gradient energy is zero for the in-plane flow of molecules
inside the membranes.  The vertical gradient energy $(\partial_z
u)^2$, on the other hand, includes the energies of the tangential
flow.  Incompressibility effects have been shown by David
\cite{incomp} and by Kleinert \cite{kleinincomp} to be irrelevant for
the renormalization of a single membrane.  Our result implies that
this is not the case for a stack of membranes.

The properties of a single membrane are obtained by letting $N \to 0$
and $ z_{\rm c} \to 0$.  In particular, the flow equation
(\ref{betaalpha}) of the ben\-ding rigidity reduces in this limit to
the known result \cite{peliti}.  It has no fixed point other than the
trivial one, which is unstable now, implying that a single membrane is
always in the crumpled strong-coupling phase.

\section{Weak-coupling ordered phase} \label{weak}

We next wish to study the low-tempe\-rature phase of the stack in more
detail, where the coupling constant $\alpha$ is weak.  As in smectic-A
liquid crystals, the theory (\ref{ourmodel}) can here safely be
approximated by taking into account only the quadratic terms.  An
important characteristic is the vertical fluctuation width or
roughness $\ell$ of a membrane in the stack defined by the mean square
{\it height} fluctuation as $\ell^2 = \langle u^2 \rangle$
\cite{lipbook}.  In the harmonic approximation, it is given by the
one-loop integral
\begin{equation}\label{rough}
\ell^2 = \langle u^2 \rangle = \frac{k_{\rm B} T}{L_{\parallel}} \sum_{n=-N/2}^{N/2}
\int_{1/L_{\perp}}^\Lambda \frac{{\rm d}^2 q_\perp}{(2 \pi)^2}
\frac{1}{B \omega_n^2 + K q_\perp^4},
\end{equation}  
where, in the absence of a surface tension, the largest wavelength is
equal to  the inverse lateral size $1/L_\perp$ of the membranes.
As first observed by Peierls and Landau, the mean square fluctuations
diverge in the infrared.  They thus destroy the long-range {\it
positional} order in the layered system at any finite temperature.  More
specifically, one finds \cite{lipbook,DGbook}
\begin{equation} \label{ell}
\ell^2 \sim \frac{k_{\rm B} T}{2 \pi} \left[\frac{L_\perp^2}{K
		L_\parallel} + \frac{1}{\sqrt{BK}} \ln(L_\perp/a) \right] .
\end{equation}  
The first contribution, only present in a finite stack, stems from the
$n=0$ term in the sum in Eq.\ (\ref{rough}).  It corresponds to a super
soft mode, where the membranes undulate coherently with constant
interlayer distance.  The second contribution, on the other hand, is
also present when the stack is infinite.  This contribution increases
slowly with the lateral size.  

Before proceeding, let us pause for a moment and consider the roughness of
the two limiting cases of our theory: a single, tensionless membrane and
an infinite, continuum stack of such membranes.  In this way, we find
\begin{equation}  \label{magda} 
\ell^2 = \left\{ \begin{array}{ll} \displaystyle \int \frac{{\rm d}^D
q_\perp}{(2 \pi)^D} \frac{k_{\rm B} T}{\kappa q_\perp^4}, & (\mbox{single
membrane}) \\ \displaystyle \int \frac{{\rm d} q_z {\rm d}^D q_\perp}{(2
\pi)^{D+1}} \frac{k_{\rm B} T}{B q_z^2 + K q_\perp^4}, & (\mbox{infinite
stack}),
\end{array} \right.
\end{equation} 
where instead of a 2-dimensional membrane we consider a $D$-dimensional
object.  It follows that for $D>D_{\rm u} = 4 - z_{\rm c}$, the roughness is
finite in the infrared, indicating that $D_{\rm u} = 4 - z_{\rm c}$ is the
upper critical dimension.  Recall that for a single membrane $z_{\rm c}=0$,
while for an infinite stack $z_{\rm c}=2$.  To determine the lower critical
dimension, we consider the mean square normal, or {\it orientational}
fluctuations $\langle (\partial_\perp u)^2 \rangle$.  This results in an
additional factor of $q_\perp^2$ in the numerator of the integrands in
(\ref{magda}).  The resulting expressions are finite in the infrared for
$D>D_{\rm l} = 2 - z_{\rm c}$, identifying $D_{\rm l}$ as the lower critical
dimension.  Hence, in going from the limit of a single, tensionless
2-dimensional ($D=2$) membrane to the opposite limit of an infinite stack,
we go from the lower critical dimension of the former to the upper critical
dimension of the latter.

Another characteristic of the weak-coupling phase of low-tempe\-rature
is the behavior of the structure factor
\begin{equation} \label{corr}
S_n({\bf x}) =\langle \exp \{i n q_0 [u({\bf x})-u(0)] \} \rangle,
\end{equation}
where $q_0$ is parallel to the $z-$axis, with $|q_0|=2\pi/l$.  This
correlation function can be directly observed in X-ray scattering
experiments, where the fluctuation spectrum is expressed as
half-widths at half-maximum of the anomalous Bragg peaks.  As in
smectic-A liquid crystals \cite{caille,ChLu}, the Fourier transform of
the structure factor has algebraic singularities at $q_z = n q_0$:
\begin{equation} \label{power} 
S_n(0,q_z) \sim (q_z - n q_0)^{-2 + n^2 \eta}, \;\;\; S(q_\perp,0) \sim
q_\perp^{-4 + 2 n^2 \eta},
\end{equation} 
with exponent $\eta$.  In the harmonic approximation, $\eta$ can be
calculated from Eq.\ (\ref{corr}) and turns out to be the same as for an
infinite stack \cite{caille}
\begin{equation}
\eta = \frac{k_{\rm B} T}{8 \pi} \frac{q_0^2}{\sqrt{B K}}.
\end{equation}
The algebraic singularities in (\ref{power}) reflect the
quasi-long-range periodic order along the stack axis.  As for smectic-A
liquid crystals, the exponent $\eta$ is temperature independent.  This
can be seen by remembering that by simple scaling arguments
\cite{helfstack}
\begin{equation}
B \sim \frac{(k_{\rm B}T )^2}{\kappa} \frac{l}{(l-w)^4},
\end{equation} 
for a stack of membranes of rigidity $\kappa$ and thickness $w$.
Specifically,
\begin{equation} 
\eta \sim \left(1 - \frac{w}{l}\right)^2.
\end{equation} 
For smectic-A liquid crystals, this expression was confirmed
experimentally \cite{exp}.

As the temperature increases, thermal fluctuations become stronger,
and eventually overcome the vertical forces of the $B$-term leading to
a vertical disordering of the stack.  In the strong-coupling
disordered phase, the Gaussian approximation used in this section to
investigate the weak-coupling ordered phase breaks down.  To study
this phase, a nonperturbative method is required.

\section{Conclusions} \label{concl}

We have shown that a stack of membranes with nonlinear curvature
energy melts at some critical temperature $T_{\rm c}$.  In the
weak-coupling low-temperature phase the system forms a periodic array
of well-defined surfaces.  There is long-range orientational order in
the pla\-nes, and quasi-long-range positional order along the stack
axis.  This phase can be accurately described by the harmonic
approximation of the Helfrich model, which coincides with the de
Gennes' theory of smectic-A liquid crystals.

Upon approaching $T_{\rm c}$, the stack melts and the lamellar phase
goes over into a strong-coupling disordered phase.  This phase cannot
be described by the harmonic approximation. Its properties will be
investigated separately in a nonperturbative framework, in the limit
of infinite embedding space dimension.

\end{document}